\documentclass[useAMS,usenatbib]{mn2e}
\usepackage[dvips]{graphicx}
\usepackage{fixltx2e}
\usepackage{epstopdf}
\usepackage{times}
\usepackage{amsmath}
\usepackage{color}
\newcommand{\aap}{A\&A}

\newcommand{\mnras}{MNRAS}

\newcommand{\apj}{ApJ}
\newcommand{\apjl}{ApJ}

\newcommand{\aj}{AJ}

\newcommand{\ncvgamma}{N_{\rm cv}-\Gamma}
\newcommand{\ncv}{N_{\rm cv}}
\newcommand{\mncvgamma}{n_{\rm cv}-\gamma}
\newcommand{\mncv}{n_{\rm cv}}
\title[CV formation in globular clusters]{Dynamical Formation of Cataclysmic Variables in Globular Clusters}
\author[J. Hong et al.]  {Jongsuk Hong$^1$, Enrico Vesperini$^1$, Diogo Belloni$^{2,3}$ \& Mirek Giersz$^2$\\
  $^1$Department of Astronomy, Indiana University, Bloomington, IN, 47401, USA\\
  $^2$Nicolaus Copernicus Astronomical Center, Bartycka 18, 00–716 Warsaw, Poland\\
  $^3$CAPES Foundation, Ministry of Education of Brazil, DF 70040-020, Brasilia, Brazil\\
}
\begin{document}

\maketitle

\label{firstpage}

\begin{abstract}
The formation and evolution of X-ray sources  in globular clusters is likely to be  affected by the cluster internal dynamics and the stellar interactions  in the cluster dense environment.
Several observational studies have revealed a correlation between the number of X-ray sources and the stellar encounter rate and provided evidence of the role of dynamics in the formation of X-ray binaries.
We have performed a survey of Monte-Carlo simulations aimed at
exploring the connection between the dynamics and formation of cataclysmic variables (CVs)  and the origin of the
observed correlation between the number of these objects, $\ncv$, and the
stellar encounter rate, $\Gamma$. 
The results of our simulations show a correlation between $\ncv$ and $\Gamma$ as found in observational data, illustrate the essential role played by dynamics, and shed light on the dynamical history behind this correlation. 
CVs in our simulations are more centrally concentrated than single
stars with masses close to those of turn-off stars, although this
trend is stronger for CVs formed from primordial binaries undergoing exchange encounters, which include a population  of more massive CVs absent in the group of CVs formed from binaries not suffering any component exchange.
\end{abstract}

\begin{keywords}
globular clusters:general.
\end{keywords}

\section{Introduction}
\label{sec:intro}
In the high-density environments of globular clusters (GCs) dynamical interactions between stars and binaries can play an important role in the formation of exotic objects such as X-ray binaries \citep[see e.g.,][]{2010AIPC.1314..135H}.
Low-mass X-ray binaries, for example, are overabundant in GCs compared to the field population \citep[see e.g.,][]{1975ApJ...199L.143C,2003ApJ...595..743S,2004ApJ...613..279J} suggesting that the cluster environment does indeed affect their formation rate.
The formation of other X-ray sources such as cataclysmic variables (CVs) and millisecond pulsars is also likely to be affected by the cluster internal dynamics and the enhanced stellar interaction rate in the dense environment of GCs.
A number of studies (see e.g. Verbunt \& Hut 1987, Pooley et al. 2003, Hui et al. 2010, Maxwell et al. 2012, Bahramian et al. 2013)  have found that the number of X-ray sources is correlated with the encounter rate of GCs and provided strong evidence that dynamical interactions do indeed play a key role in the formation of X-ray binaries.

The mechanisms that have been suggested for the formation of X-ray binaries include: (1) binary stellar evolution of primordial binaries,
(2) binary interactions affecting the binary orbital parameters and exchange interactions \citep[interactions during which one of the initial binary components is replaced by the interacting single star or one of the components of the other interacting binary; during such interactions a binary can acquire a dark remnant component;][]{1976MNRAS.175P...1H}, (3) three-body binary formation \citep{1975ApJ...199L.143C}, (4) tidal captures \citep{1975MNRAS.172P..15F}.

Although there is a vast literature on the stellar evolution of
binary stars, few numerical studies have attempted to explore the
formation of X-ray sources in the context of models including the effects of the dynamical evolution of globular clusters (see e.g. an early study by Davies 1997 and the more recent studies by Ivanova et al. 2006, Shara \& Hurley 2006); the development of codes capable of 
leveraging the increasing computational power available is now opening the possibility  to
carry out detailed numerical simulations following a cluster dynamical evolution and the effects of dynamics on the cluster stellar populations.

In this study, we have carried out a survey of Monte-Carlo simulations
following the dynamical evolution of clusters with a variety of
different initial conditions and explored the formation of CVs and the
interplay between CV formation and the dynamical evolution of the
cluster structural properties.
The outline of this paper is the following:
in Section 2, we present our methods and initial conditions; in Section 3 we present the results of our simulations; we
summarize our results in Section 4. 

\section{Methods and Initial Conditions}

For our simulations we have used the MOnte-Carlo Cluster simulAtor ({\sc mocca}) code developed by Giersz et al. (2008) with an implementation of the {\sc fewbody} interaction scheme (Fregeau et al. 2004) by Hypki \& Giersz (2013).

For all our simulations we have adopted a Kroupa (2001) stellar initial mass function with stellar masses ranging from 0.1 to 100 $M_{\odot}$ and metallicity $Z=0.001$.
All our simulations include a population of primordial binaries; 
 the initial binary distribution follows the Kroupa's (1995; 2013) description,
based on the eigenevolution procedure that converts a birth population to an initial population.
The effects of both single stellar evolution (SSE; Hurley et al. 2000) and binary stellar evolution (BSE; Hurley et al. 2002) are  implemented in the MOCCA code and included in our simulations.

We have explored clusters with different initial masses, half-mass
radii, galactocentric distances, and primordial binary fraction (larger number of soft binaries could be present but would be rapidly destroyed in these dense clusters); the values explored are summarized in Table 1. We have run simulations for all the possible combinations of the
different values reported in Table 1 for a total number of 81
simulations. For all the systems, the initial density profile is  that
of a King (1966) model with a value of the central dimensionless
potential, $W_0$,  equal to 7. 

In all the simulations the effects of a tidal cut-off are included and determined  according to the cluster mass and galactocentric distance (see e.g. Giersz et al. 2013). For the initial conditions considered in our survey the ratio of the half-mass radius to the tidal radius spans a range between about 0.005 and  0.09.

In the analysis of our simulations, we identify a binary system as a
CV if it is composed of a white dwarf (WD) accreting mass from a companion main-sequence star filling its Roche lobe.  
Many studies attempting to establish a connection between the number
of  X-ray sources and the dynamics of globular clusters have focused
their attention on the stellar encounter rate, $\Gamma$, as a
parameter providing  a quantitative measure of the role of the cluster
dense environment in the formation of these sources.
$\Gamma$ is calculated in our analysis as the volume integral over the
entire cluster of $\rho^2/\sigma$, where $\rho$ is the density and
$\sigma$ is the 1-D velocity dispersion (see e.g. Pooley et al. 2003). 
\begin{table}
  \begin{center}
  \caption{Initial parameters}
  \begin{tabular}{l c }
  \\
    \hline
    \hline
    Properties & Variation \\
    \hline
    $N$ & $2\times10^5, 5\times10^5,  10^6$\\
    $R_{\rm g}$ (kpc) & 4, 8, 16  \\
    $r_{\rm h}$ (pc) & 1, 2, 4  \\
    $f_{\rm b}$ & 10\%, 20\%, 50\% \\
    \hline
  \end{tabular}
  \end{center}
\begin{flushleft} 
 $N=N_s+N_b$ is the sum of the number of single, $N_s$, and binary, $N_b$, particles. $N_{\rm tot}=N_s+2N_b$ is the total number particles; the initial mean stellar mass is $\langle m \rangle=M/N_{\rm tot} \simeq 0.65 M_{\odot}$ (where $M$ is the initial cluster mass)\\
$R_{\rm g}$ is the galactocentric distance.\\
$r_{\rm h}$ is the initial half-mass radius of clusters.\\
$f_{\rm b}=N_b/(N_s+N_b)$ is the initial binary fraction.\\
\end{flushleft}
\end{table}
\section{Results}

\begin{figure}
  \includegraphics[width=84mm]{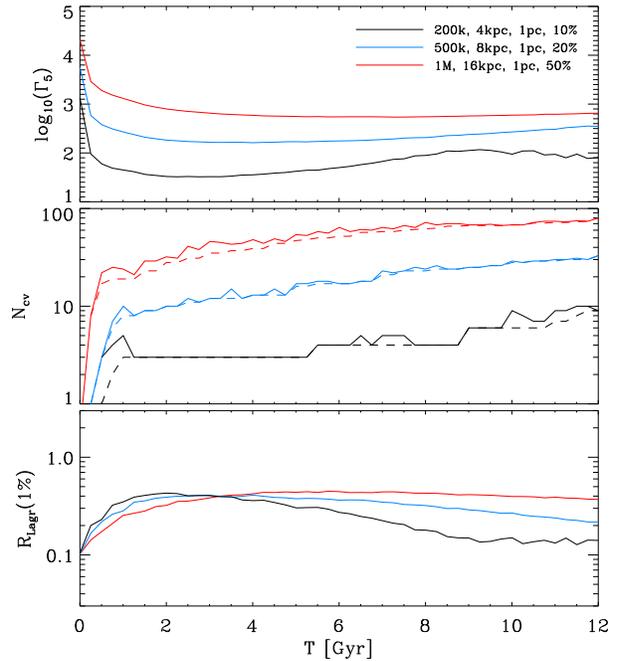}
  \caption{Time evolution of stellar encounter rates $\Gamma_5$ ($\Gamma_5=\Gamma/10^5$) (top panel), number of CVs (middle panel) and 1 per cent Lagrangian radius (bottom panel) for three simulations from our survey. The legend in the top panel indicates the initial values of $N_{\rm tot}, R_{\rm g}, r_{\rm h}, f_{\rm b}$ for the three models.  In the middle panel the dashed lines show the time evolution of $\ncv$ including only the CVs which are part of the CV population at 12 Gyr.}
\end{figure}

We start the presentation of our results by focusing our attention on
the time evolution of the encounter rate, $\Gamma$, shown in the top panel of Figure 1 for a few representative models (notice that in all the figures, we show the value of $\Gamma_5=\Gamma/10^5$). 
All our models are initially compact and characterized by large
initial values of $\Gamma$. As the clusters expand in response to 
mass loss due to stellar evolution, $\Gamma$  decreases rapidly with
time; this early evolution is followed by a phase during which, at first,
$\Gamma$ does not vary significantly  and then slightly increases as
the system evolves towards higher central densities (and eventually
core collapse);  it is interesting to notice that
once the cluster enters the phase in
which core contraction is halted and supported by the energy provided by primordial
binaries, $\Gamma$ stops growing again as shown by the time evolution
of $\Gamma$ for the model 200k particles (black line in Figure 1). The middle panel of
Figure 1 shows, for the same set of representative models, the time evolution
of the number of CVs, $\ncv$. A comparison of the top and middle panels
clearly shows the link between the evolution of $\ncv$ and
$\Gamma$. For all the models, $\ncv$ is characterized by a rapid early
growth (see also Belloni et al. 2016b) followed by milder increase during the rest of the cluster evolution. 

Finally, in order to better illustrate the connection between the
cluster inner structural evolution and the evolution of $\Gamma$ and $\ncv$
we plot in the bottom panel of Figure 1 the time evolution of the 1
per cent Lagrangian radius for the same models considered in the
previous panels. The time evolution of this Lagrangian radius allows us
to clearly identify the different evolutionary stages described above:  the early expansion driven by mass loss due to stellar evolution followed by a gradual contraction (here visible for the models with 200k and 500k particles) and, for the model with 200k particles, the phase during which core collapse is halted by the energy provided by binaries.

In Figure 2, we further illustrate the connection between $\ncv$ and
$\Gamma$ by showing the number of CVs at 12 Gyr for all our models as a
function of the encounter rate measured at 12 Gyr. 
The models not shown in this
figure undergo complete dissolution before 12 Gyr.   
The number of CVs in a system  depends on the initial cluster
mass, the initial half-mass radius and the initial binary
fraction. 
The results of our simulations show a clear trend for  $\ncv$ to
increase with $\Gamma$ consistent, in general, with what found in the
observational studies discussed in Section 1; the dashed line plotted in Figure 2 is not a fit to the data but a line with a slope in the ($\log \ncv-\log \Gamma$) plane equal to that found in observational data (Verbunt 2007).

The inset panel of Figure 2 shows the evolutionary tracks in the
$\ncvgamma$ plane for the few representative models already selected
for Figure 1 and  illustrate the dynamical history behind the trend between the values of $\ncv$ and $\Gamma$ measured at 12 Gyr. As already illustrated in Figure 1, our models reach their final (at 12 Gyr) position in the $\ncvgamma$ plane moving, in general, from initially larger values of $\Gamma$ and increasing $\ncv$ as they evolve.

Figure 3a shows a plot of the number of CVs versus the cluster encounter rate (both measured at 12 Gyr) in which both these quantities are normalized to the cluster mass at 12 Gyr: the mass-normalized quantities are defined, respectively, as $n_{\rm cv} \equiv N_{\rm cv}/(M/10^6M_{\odot})$ and $\gamma \equiv \Gamma_5/(M/10^6M_{\odot})$.
The  $\mncvgamma$ plot has been introduced by  
Pooley \& Hut (2006) to take into consideration the observed trend between the encounter rate and the cluster mass. In that study, Pooley \& Hut (2006)  fit the observed trends between these two quantities for different classes of X-ray sources with a function   $\mncv=a\gamma^{b}+c$.
In this figure, our simulation results show a larger spread compared to the absolute number and encounter rate found in Figure 2; the dashed line in Figure 3a shows the result of fitting all the results of our simulations with a function equal to that adopted by Pooley \& Hut (2006) to fit the available observational data; the specific values of the parameters of our best fit are $(b, c)=(0.73\pm0.14, 41.32\pm5.63)$.  The power-law index $b$  we find here is consistent with that found by Pooley \& Hut (2006) for the class of X-ray sources they consider to be dominated by CVs (class II in their paper).
Figure 3b shows the evolutionary tracks on this plane of a
few representative models and illustrates the dynamical history behind
this trend: as shown also in Figure 2, clusters, in general, evolve
towards their final (at 12 Gyr) position on this plane from larger
initial values of the normalized  encounter rate. For some clusters,
the normalized encounter rate after an early decrease can increase
again as the cluster evolves towards core collapse and loses mass.

\begin{figure}
  \includegraphics[width=84mm,trim={5mm 5mm 5mm 5mm}]{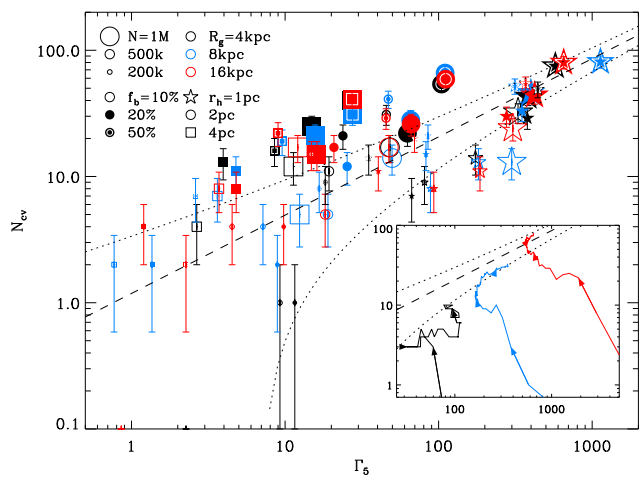}
  \caption{Number of CVs, $\ncv$, at 12 Gyr as a function of the cluster encounter rate ($\Gamma_5=\Gamma/10^5$). 
  The error bar represent the Poisson error of $\ncv$.
  Different symbols in size, color, type and shape show different number, galactocentric distance, binary fraction and half-mass radius, respectively.
  Dashed line shows the observational trend with a slope of 0.62 (Verbunt 2007). 
  Upper and lower dotted lines are twice of the Poisson error of the line of the observational trend.
  The inset figure shows the evolutionary trajectories on the $\ncv-\Gamma_5$ plane for the three selected models shown also in Figure 1.  Three arrows (at $t=0.5,3,10$ Gyr) have been plotted on each line to indicate the direction of evolution.}
\end{figure}

\begin{figure*}
  \includegraphics[width=160mm]{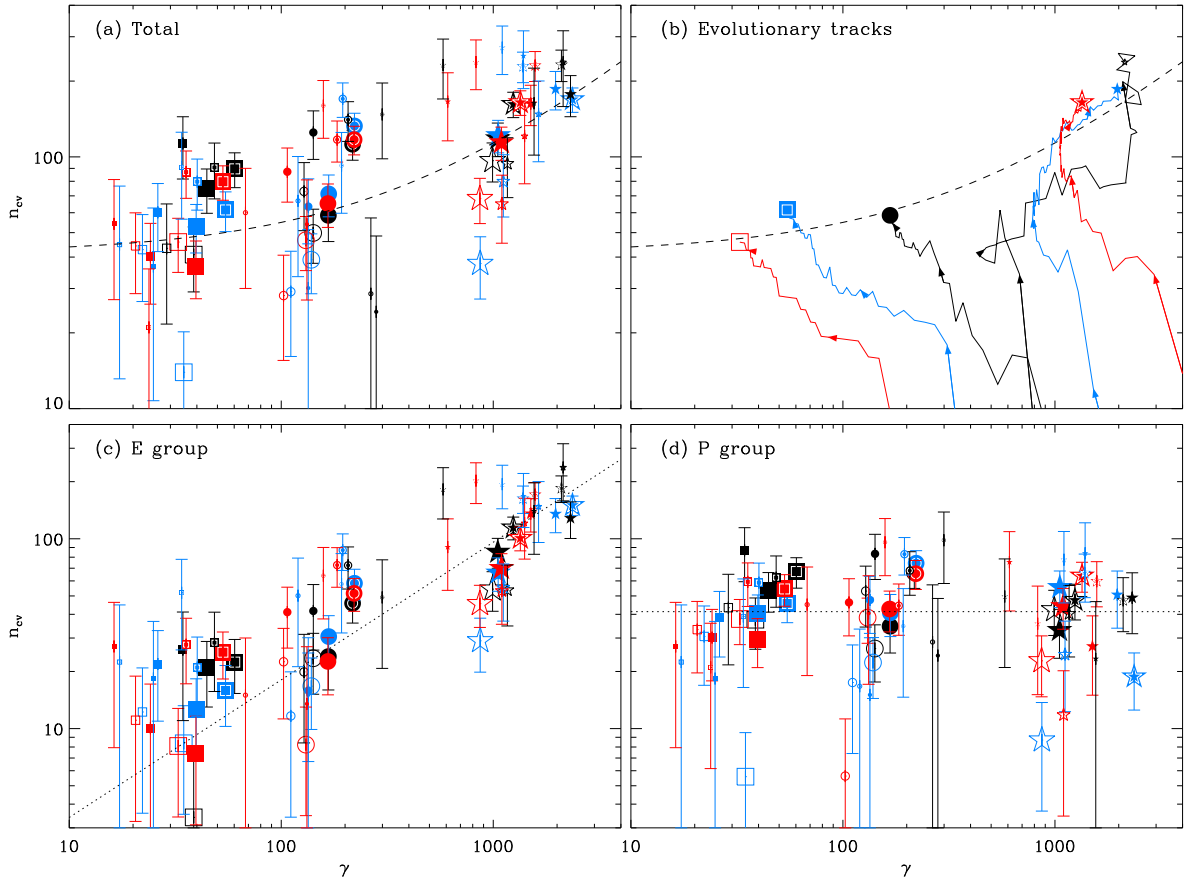}
  \caption{{\it Top left-hand panel:} mass-normalized number of CVs $n_{\rm cv}\equiv N_{\rm CV}/(M/10^6M_{\odot})$ at 12 Gyr as a function of the mass-normalized encounter rates $\gamma\equiv\Gamma_{\rm 5}/(M/10^6M_{\odot})$ (where $M$ is the cluster mass at 12 Gyr). Symbols are the same as Figure 2. The dashed line is the result of fitting of our data (see section 3). 
{\it Top right-hand panel:} Evolutionary tracks for some simulations in $\mncvgamma$ plane.  Three arrows (at $t=0.5,3,10$ Gyr) have been plotted on each line to indicate the direction of evolution. 
{\it Bottom left-hand panel:} $\mncv$ vs $\gamma$  for CVs in E group. A dotted line with the same value of the power-law index $b$ (see section 3) obtained in the fit of Figure 3a is also shown.  
{\it Bottom right-hand panel:} $\mncv$ vs $\gamma$ for CVs in P group. A dotted line corresponding to a constant value equal to the $c$ parameter (see section 3) obtained in the fit of Figure 3a is also shown.}
\end{figure*}
\begin{figure}
  \includegraphics[width=84mm]{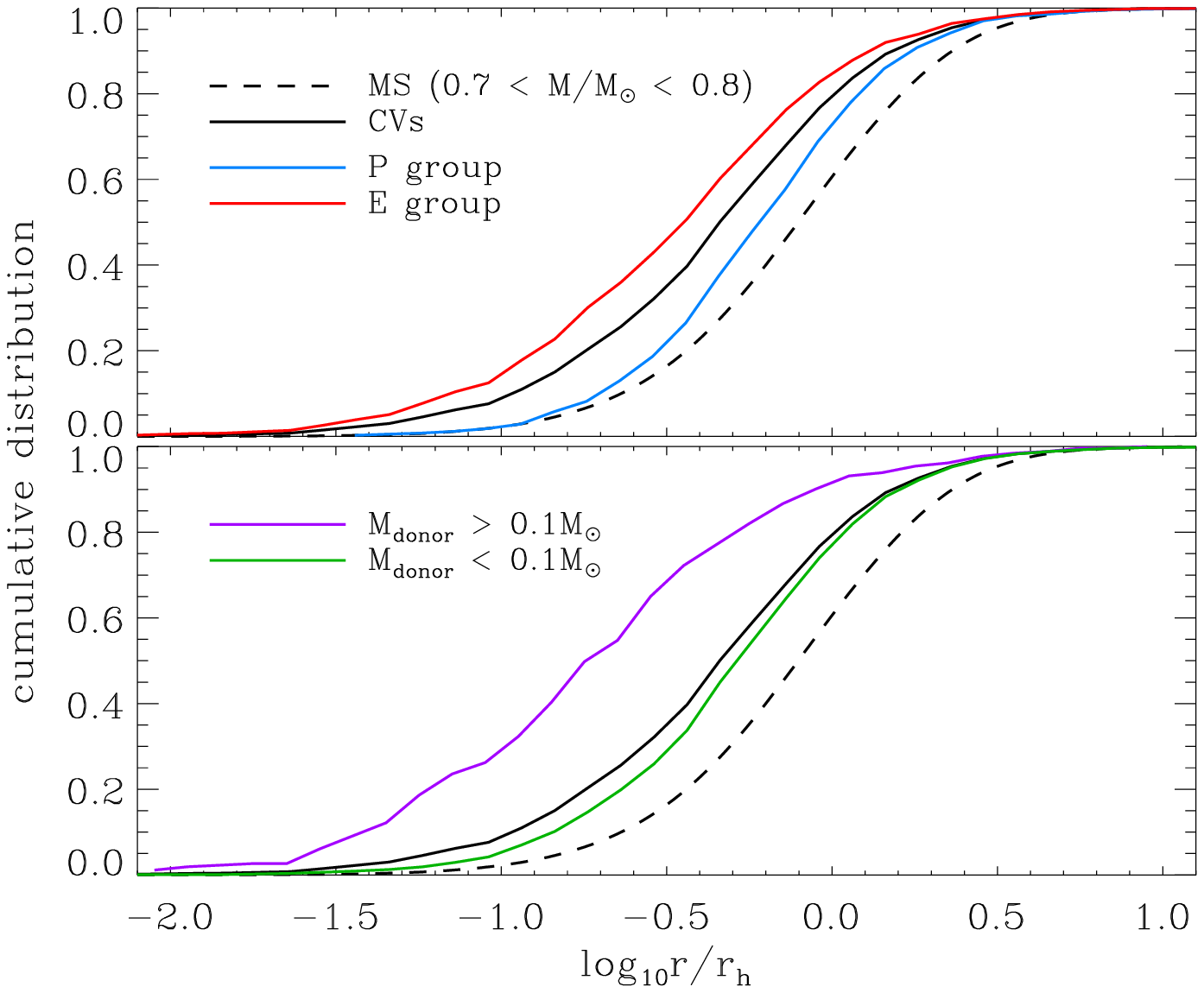}
  \caption{Cumulative radial distribution of CVs (solid lines) and main sequence stars with the mass between 0.7 and 0.8$M_{\odot}$ (dashed line) from all of our models. The radial distance of each CV is normalized to the cluster half-mass radius. Blue and red lines in the top panel show, respectively, the distribution of CVs in the P group (blue line) and the E group (red line) (see text for more details on the definition of these groups). In the bottom panel the purple (green) line shows the distribution of CVs with a main sequence donor more (less) massive  than 0.1 $M_{\odot}$. The black line in each panel shows the radial distribution of the whole CV population.}
\end{figure}

Essentially all the  progenitors of the CVs formed in our simulations
are from the population of primordial binaries.  In agreement with the
results found by Belloni et al. (2016a) for a population of binaries with the same Kroupa (1995, 2013) initial conditions adopted here, we find that for  all of the CVs dynamics plays a key role in their evolution towards becoming a
CV (see Belloni et al. 2016a for further discussion). 
 We can, however, distinguish two groups in the CV population: one
group consisting of CVs  formed  
from primordial binaries which during the cluster evolution had their
orbital parameters affected by encounters with other binaries and
single stars, but their components at 12 Gyr are still the initial
ones (hereafter we refer to this group as the P(rimordial) group), and
another group consisting of CVs formed instead from primordial binaries which suffered at least one exchange
encounter during which one of the initial components was replaced by
either the interacting single star or one of the components of the
interacting binary (we refer to this group as the E(xchange) group). 
Considering all the CVs in our models at 12 Gyr, the ratio of the number of CVs in the E group to those in the P group is equal to about 1.19.

Figures 3c and 3d show  the $\mncvgamma$ plot
separately for CVs in the E and P group, respectively. These
plots show a a clear difference in the trend of the mass-normalized
number of CVs versus the mass-normalized encounter rate. As already pointed out above,  dynamics plays a key role for both the E and the P groups; however, these two
plots shed further light on the link between the origin of CVs
belonging to these two groups and the cluster environment; $\Gamma$ measures the
importance of close encounters which can lead to an exchange in the
binary component and produce CVs in the E group. CVs in the P group, on the other hand, are the
result of orbital perturbations of primordial binaries and, for all
the initial conditions explored in this paper, are
produced at a rate independent of $\Gamma$.  The spread observed in Figure 3c and 3d appears to be mainly determined by the binary fraction as systems with a larger binary fraction tend, in general, to produce more CVs. 
The plateau in $\mncv$ for small values of $\gamma$ shown in Figure 3a and found in observations is, in the context of our models, the consequence of the fact that CVs in the P group are produced with the same efficiency independently of $\gamma$; this parameter is instead more directly linked to the formation of CVs in the E group and is responsible for the increase of $\mncv$ for systems with larger values of $\gamma$.
 As shown by Figures 3c and 3d the ratio of the number of CVs in the E group to the number of CVs in the P group increases for more compact systems characterized by larger values of $\gamma$.

We conclude with a discussion of the radial distribution of CVs. In Figure 4 we show the cumulative radial distribution of CVs at 12
Gyr for all the CVs in all of our models and for CVs of the E and the P
group separately (top panel). In the bottom panel of Figure 4 we also divide the CV population according to whether the mass of the main sequence component is more or less massive than 0.1 $M_{\odot}$.
We compare the radial distribution of CVs
with that of single main sequence stars with masses between 0.7 and
0.8 $M_{\odot}$. CVs are, in general,  more centrally concentrated than the 
main sequence stars: this is due to the presence of CVs currently more massive than the main sequence stars considered as well as to the remaining memory of the CV formation history and progenitor masses.

As shown in Figure 4 the difference in the radial distribution with
main sequence stars is stronger for CVs in the E group and for CVs with main sequence component more massive than 0.1 $M_{\odot}$; this is a
consequence of the fact that these groups include a population of more massive CVs as shown in Figure 5 where we have plotted the mass distributions of CVs in the different groups.

The trend in the spatial distribution found in our simulations is consistent with the results of the observational study
of Cohn et al. (2010), who
have compared the radial distribution of CVs with that of Main
Sequence Turn-off stars in NGC6397 and found the CVs (in particular the population of bright CVs) to be more centrally concentrated. A similar trend in the spatial distribution of CVs has also been found in a recent study of NGC 6752 (Lugger et al. 2016, in prep.).

We finally point out that the presence of a tail of
more massive CVs in the E group is consistent with the results of
Shara \& Hurley (2006) and Belloni et al. (2016a) who also found that
the population of CVs affected by exchange encounters included a population of more massive CVs. CVs with main sequence donor larger than 0.1 $M_{\odot}$ belong mainly to the E group and have, in general, younger ages than those with a donor less massive than 0.1 $M_{\odot}$.

\begin{figure}
  \includegraphics[width=84mm]{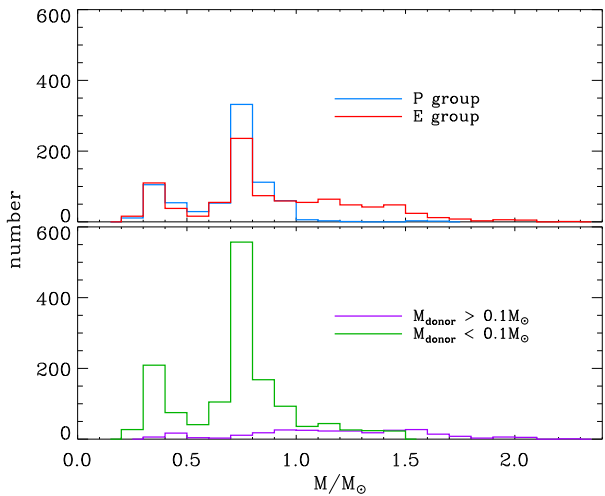}
  \caption{Distribution of the total CV mass  at 12 Gyr for  all the CVs in all our simulations. The top panel show the mass distribution for CVs in the E and the P group separately; the bottom panel shows the mas distribution of CVs with a main sequence donor more or less massive than 0.1 $M_{\odot}$.}
\end{figure}

\section{Summary}
In this paper we have presented the results of a survey of Monte Carlo simulations of globular cluster dynamical evolution  aimed at exploring the connection between a cluster dynamical evolution and the formation of CVs.
Dynamical effects in the dense stellar environment of globular clusters can play a key role in the formation and evolution of a variety of stellar populations and one of the key observational signatures  of this link is the correlation between the number of X-ray sources and the stellar encounter rate. 

In our survey we have explored the evolution of clusters 
with a variety of different initial conditions spanning different values of the cluster mass, size, binary fraction, and galactocentric distances and followed the evolution of the number of CVs along with that of the cluster encounter rate, $\Gamma$.

After 12 Gyr of dynamical evolution our models show a correlation
between the number of CVs and $\Gamma$ consistent with that found in
observational studies. Our simulations show how such a
correlation arises and the dynamical history behind it (Figure 2 and
Figure 3a and 3b). The
clusters we have considered are initially compact and characterized by
large values of $\Gamma$ which  decrease during the cluster early
evolution as the cluster expands in response to mass loss due to
stellar evolution. This early evolutionary phase is followed by the
cluster long-term evolution during which the encounter rate is 
approximately constant at first and then slightly increases again as
the cluster approaches core collapse. The number of CVs increases
rapidly during the cluster early evolution and, after this early phase,
is characterized by a milder growth.

The progenitors of the CVs are essentially all from the population of
primordial binaries; in all cases, dynamics plays an essential role in
the evolution towards becoming a CV but  we have identified two groups
of CVs: one group (the P group) formed from primordial binaries  whose
orbital parameters are affected by encounters but whose initial
components are the same during the entire cluster evolution and
another group (the E group)  from primordial binaries which undergo
at least one exchange encounter during which one of the initial
components is replaced by one of the interacting stars. The
mass-normalized number of CVs in the E group is found to increase with
the mass-normalized encounter rate (Figure 3c), while, for the range of
initial conditions explored in this paper, the number of CVs
in the P group is approximately independent of the encounter rate
(Figure 3d). The CVs in the E group are therefore those responsible for the observed increase in the mass normalized $n_{CV}$ with the mass normalized encounter rate.

Consistently with the results of observational studies, we find that CVs tend to be more centrally concentrated than
single main sequence stars with masses close to the turn-off stars at
12 Gyr. This trend is found to be stronger for CVs in the E group which include a populations of more massive CVs not found in the P group; we have also divided the CV population according to the mass of the main-sequence donor: CVs with a main sequence component more massive than 0.1 $M_{\odot}$ belong mainly to the E group and they tend to have younger ages and be more concentrated than those with a main sequence component less massive than 0.1 $M_{\odot}$. 
 In future studies we will further expand the analysis presented here to study the evolution of different X-ray sources and a broader range of initial conditions including larger fractions of primordial binaries (see e.g. Marks \& Kroupa 2012) and different cluster structural parameters.

\section*{Acknowledgements}

This research was supported in part by Lilly Endowment, Inc., through its support for the Indiana University Pervasive Technology Institute, and in part by the Indiana METACyt Initiative. The Indiana METACyt Initiative at IU is also supported in part by Lilly Endowment, Inc.
DB and MG were partially supported by the Polish National Science Centre (PNSC) through the grant DEC-2012/07/B/ST9/04412.
DB was supported by the CAPES foundation, Brazilian Ministry of Education through the grant BEX 13514/13-0.

\end{document}